\documentclass{article}

\usepackage[usenames,dvipsnames,svgnames]{xcolor}

\usepackage{hyperref}
\hypersetup{
    colorlinks=true,
    citecolor=MidnightBlue,
    urlcolor=Bittersweet,
}

\usepackage[margin=1in]{geometry}

\usepackage{float}
\usepackage{bbm}
\usepackage{bm}
\usepackage[normalem]{ulem}
\usepackage{extarrows}
\usepackage{enumitem}
\usepackage{comment}
\usepackage{amsthm,amsmath,amsfonts,amssymb}
\usepackage{booktabs}
\usepackage{algorithm,algpseudocode} 

\usepackage{cleveref}

\usepackage{graphicx}

\usepackage{algorithm}
\usepackage{algorithmicx}
\usepackage{algpseudocode}
\usepackage{url}
\usepackage{verbatim}
\usepackage{makecell}
\usepackage{tikz}
\usepackage{cite}
\usetikzlibrary{spy}
\usepackage{multirow}

\usepackage{subcaption}

\newcommand{\noise}{\mathbf{n}}
\newcommand{\direct}{\mathbf{d}}

\newcommand{\scatter}{\mathbf{s}}
\newcommand{\abel}{\mathcal{A}}

\newcommand{\density}{\boldsymbol{\rho}}
\newcommand{\m}{\mathbf{m}}
\newcommand{\E}{\mathbb{E}}
\newcommand{\x}{\mathbf{x}}
\newcommand{\y}{\mathbf{y}}
\newcommand{\R}{\mathbb{R}}
\newcommand*\samethanks[1][\value{footnote}]{\footnotemark[#1]}

\begin{document}

\title{Physics-Driven Learning of Wasserstein GAN for Density Reconstruction in Dynamic Tomography
\footnote{\href{https://doi.org/10.1364/AO.446188}{© 2022 Optica Publishing Group.} One print or electronic copy may be made for personal use only. Systematic reproduction and distribution, duplication of any material in this paper for a fee or for commercial purposes, or modifications of the content of this paper are prohibited.}}

\author{
Zhishen Huang\thanks{Department of Computational Mathematics, Science and Engineering, Michigan State University, East Lansing, MI 48824, USA, huangz78@msu.edu} , 
Marc Klasky\thanks{Theoretical Division, Los Alamos National Laboratory, Los Alamos, NM, 87545, USA, mklasky@lanl.gov} ,
Trevor Wilcox\thanks{Theoretical Design Division, Los Alamos National Laboratory, Los Alamos, NM, 87545, USA, wilcox@lanl.gov} , 
Saiprasad Ravishankar\samethanks[1] \thanks{Department of Biomedical Engineering, Michigan State University, East Lansing, MI 48824, USA, ravisha3@msu.edu}
}
\maketitle

\begin{abstract}
Object density reconstruction from projections containing scattered radiation and noise is of critical importance in many applications.
Existing scatter correction and density reconstruction methods may not provide the high accuracy needed in many applications and can break down in the presence of unmodeled or anomalous scatter and other experimental artifacts.
Incorporating machine-learned models could prove beneficial for accurate density reconstruction particularly in dynamic imaging, where the time-evolution of the density fields could be captured by partial differential equations or by learning from hydrodynamics simulations. In this work, we demonstrate the ability of learned deep neural networks to perform artifact removal in noisy density reconstructions, where the noise is imperfectly characterized. We use a Wasserstein generative adversarial network (WGAN), where the generator serves as a denoiser that removes artifacts in densities obtained from traditional reconstruction algorithms. We train the networks from large density time-series datasets, with noise simulated according to parametric random distributions that may mimic noise in experiments. The WGAN is trained with noisy density frames as generator inputs, to match the generator outputs to the distribution of clean densities (time-series) from simulations. A supervised loss is also included in the training, which leads to improved density restoration performance. In addition, we employ physics-based constraints such as mass conservation during network training and application to further enable highly accurate density reconstructions. Our preliminary numerical results show that the models trained in our frameworks can remove significant portions of unknown noise in density time-series data. 
\end{abstract}

\section{Introduction}

The reconstruction of a density object from line integrated radiographic projections has a long history dating back to Radon~\cite{radon}.  In many scientific applications arising in material science, shock physics, inertial confinement fusion, and in nuclear security applications including stockpile stewardship, a sequence of radiographic images are acquired and utilized in an attempt to elucidate the physics models and their associated parameters. However, density reconstructions performed using forward modeling approaches from experimental radiographic data of dynamic tests are complicated by the noisy and complex multi-scale and multi-physics environment. The uncertainties in the reconstruction process arise from our inability to exactly represent various aspects of the radiographic measurement system such as scatter, beam spot movement, beam-target interactions, beam dynamics repeatability, and aspects of the image formation process in the forward model.

\begin{figure}[!h]
  \centering
  \begin{tabular}{cc}
\hspace{-0.2in}  \includegraphics[width=0.6\textwidth]{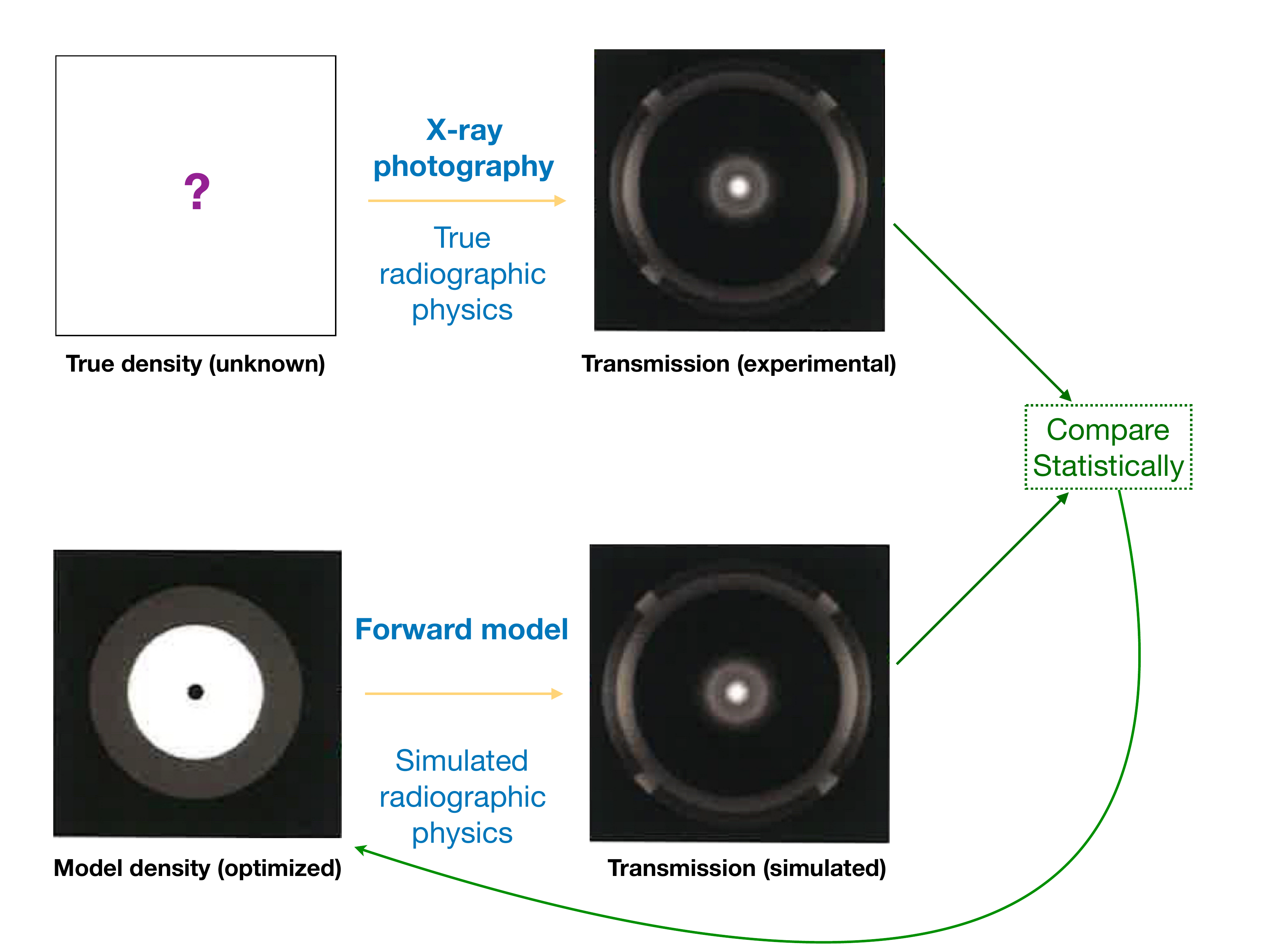} &
   \includegraphics[width=0.35\textwidth]{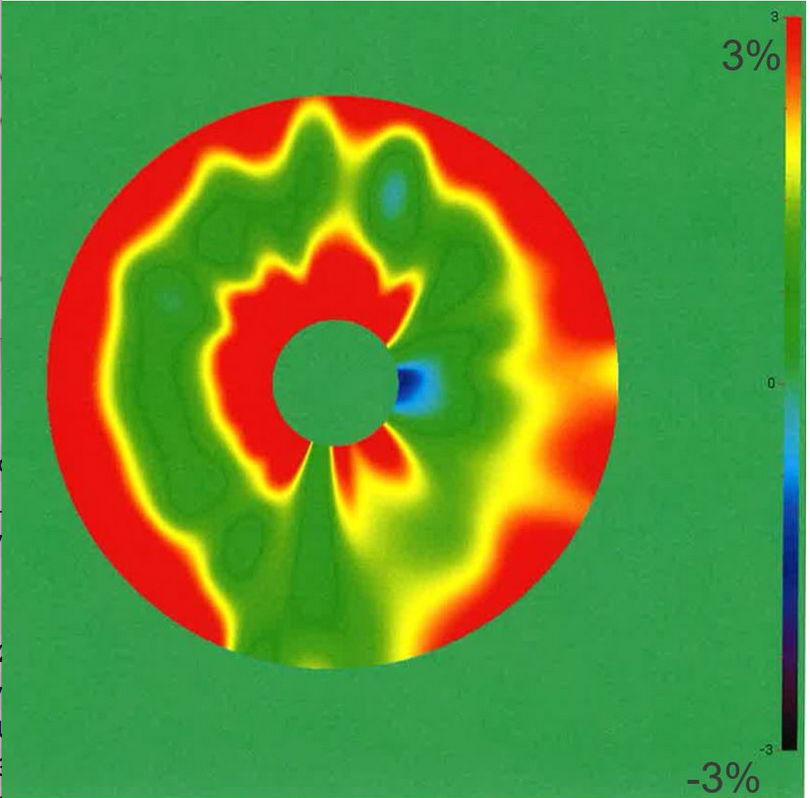}\\
   \hspace{-0.2in}  (a) & (b) \\
  \end{tabular}
  \caption{(a) Forward modeling approach to radiographic reconstruction. (b) Density errors of the French Test Object (FTO) using a forward model.}
  \label{fig:Forward}
\end{figure}

Modern approaches to solving the radiographic inversion problem for poly-energetic radiographic systems work with complex non-linear or non-convex forward models, and employ iterative reconstruction techniques~\cite{elbakri_statistical_2002}.  Mathematically, we may express the inversion problem as
\begin{equation}
\widehat{\x} = \mathrm{arg} \min_\x F(\x) + \sum_i \alpha_i R_i(\x)
\label{eq:opt_recon}
\end{equation}
where $F(\x)$ is a data-fidelity term capturing the forward model of the imaging process and the statistical models of measurements and noise, and the functions $R_i(\x)$ denote different regularizers enforcing assumed properties on the reconstructed image $\x$.  The scalar parameters $\alpha_i \geq 0$ control the relative strength of the respective regularizers.  
The determination of the correct density distribution from~\eqref{eq:opt_recon} is contingent not only upon formulating an accurate physics-based forward model and noise model, but also on imposing appropriate regularization and effective schemes to accurately solve the optimization problem. 
Figure~\ref{fig:Forward}(a) depicts the objective of the forward modeling approach.

\begin{figure}[!h]
    \centering
    \begin{subfigure}[b]{.65\textwidth}
        \centering
        \includegraphics[width=\textwidth]{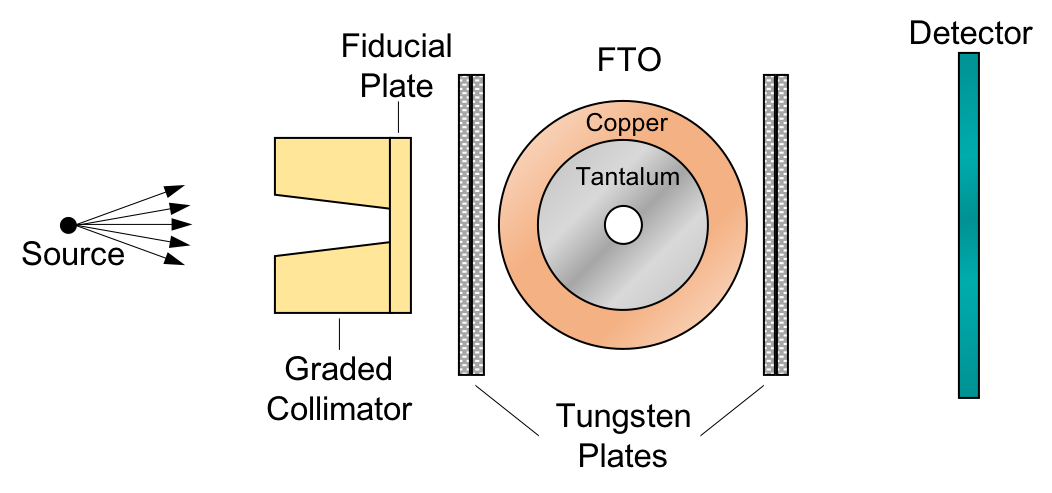}
        \label{fig:LOS}
        \caption{A schematic of the DARHT radiographic scene.}
    \end{subfigure}%
    \begin{subfigure}[b]{.35\textwidth}
        \centering
        \includegraphics[width=\textwidth]{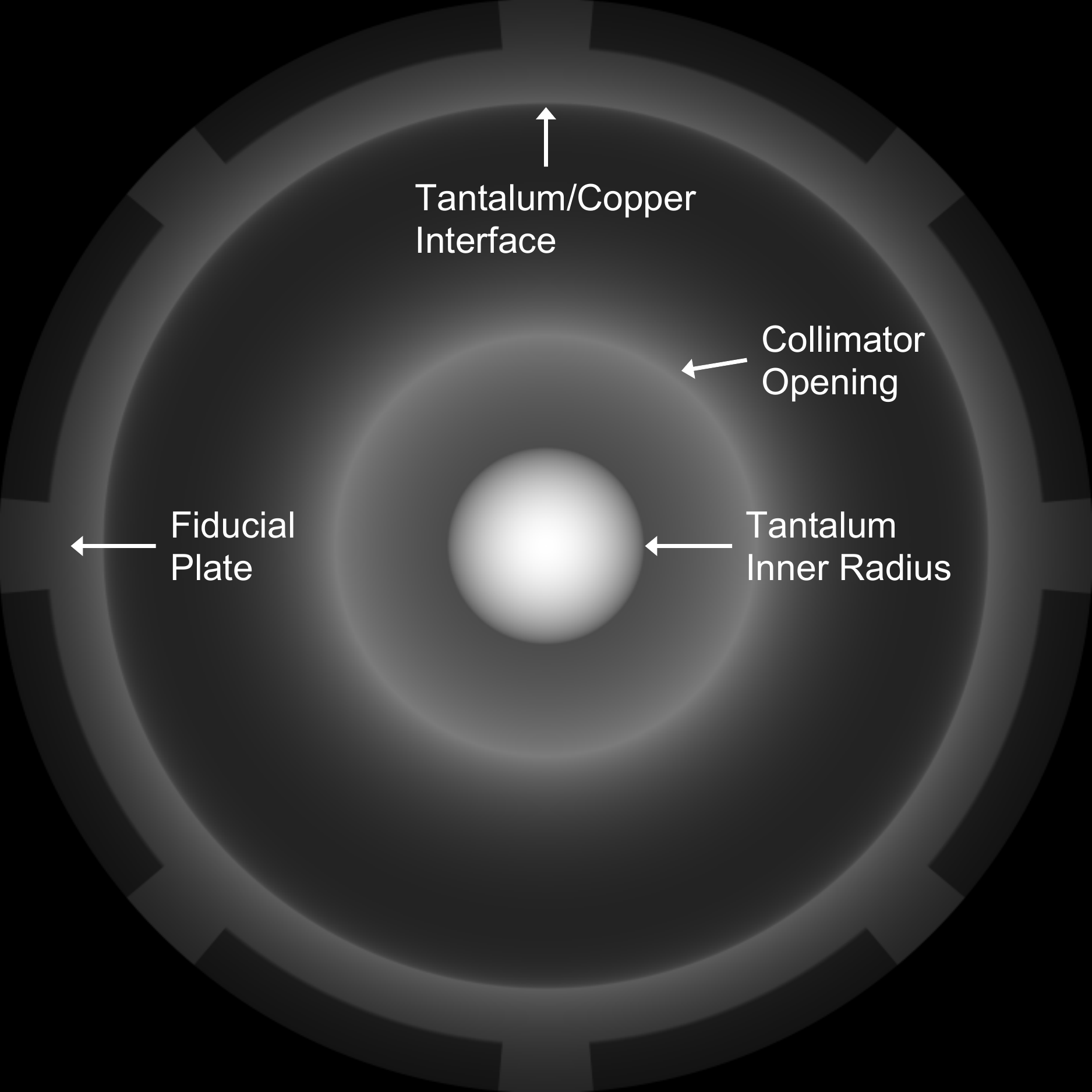}
        \label{fig:FTO_annotated}
        \caption{An FTO simulated radiograph.}
    \end{subfigure}
   \caption{The French Test Object (FTO) fielded at DARHT and a simulated radiograph with key features identified.}
  \label{fig:FTO}
\end{figure}

The practical application of the forward modeling approach to a French Test Object (FTO) fielded at the Los Alamos Dual Axis Radiographic Hydrodynamic Test Facility (DARHT), Figure~\ref{fig:FTO}, has enabled density reconstructions to be obtained.  The density errors using a radiographic forward model are depicted in Figure~\ref{fig:Forward}(b)~\cite{LA-UR-20-29499}.


While the density reconstructions performed using the traditional radiographic iterative reconstruction approach would appear to be reasonable, they are not in general sufficient to elucidate physics of many of the fundamental issues, e.g., equation of state modeling and constitutive properties of the materials. 
A major source of error in the density reconstructions is scattered radiation.
Scatter typically creates a loss of contrast and leads to image artifacts such as cupping, shading, streaks, etc.
Scatter is caused by many types of photon-matter interactions~\cite{cohentannoudji_atom_1998} including Compton scatter, Rayleigh scatter, pair production, scatter involving the scene/background, etc.
Many techniques have been proposed for scatter correction~\cite{stonestrom_scatter_1976,sun_improved_2010,bhatia_convolution_2017,tisseur_evaluation_2018,maier_deep_2018,mccann_local_2021} (cf.~\cite{ruehrnschopf_general_2011,ruehrnschopf_general_2011a} for detailed review) in medical imaging, nondestructive testing, and other settings. 
The more recent techniques~\cite{maier_deep_2018,mccann_local_2021}
learn or fit scatter models based on training data including those generated from Monte Carlo N-particle transport code (MCNP) simulations~\cite{werner_mcnp6.2_2018}.

The complex nature of scatter makes it impossible to accurately capture it in MCNP simulations. This is even more a concern for anomalous scatter fields present in experimental data such as from scene scatter.
Such fields may not be accurately characterized and removed.
Moreover, existing scatter (or other noise) correction techniques are not yet able to provide highly accurate quantitative density reconstructions~\cite{mccann_local_2021} and breakdown in the presence of anomalous scatter fields.
Thus, incorporating learned models of the underlying objects is of much importance for accurate density reconstruction. This may be particularly useful in dynamic imaging, where the time-evolution of the underlying density fields could be captured by learning from large physics-based hydrodynamics simulations, wherein various additional physics-based properties could also be incorporated during learning.

In this work, we demonstrate the ability of a learned hybrid model which leverages Wasserstein generative adversarial network (WGAN) paradigm and supervised learning to perform artifact removal in noisy density reconstructions, where the noise is imperfectly characterized. The generator in the WGAN serves as a denoiser that removes artifacts in densities obtained from traditional 
density reconstruction 
algorithms. We train the WGAN from large density time-series datasets obtained from hydrodynamic simulations, with noise simulated according to parametric random distributions that mimic noise in experiments. Essentially, the GAN is trained with noisy density frames as generator inputs, with the goal of matching the generator output distribution to the distribution of clean densities (time-series) from the hydrodynamic simulations. 
Our numerical results show that the denoisers trained in the WGAN framework are capable of removing significant portions of existing unknown noise in density time-series.
In addition, we have found that the use of physics-based constraints and regularization during network training and application allows for more accurate density reconstructions that are more consistent with the underlying hydrodynamics. In particular, we leverage mass conservation over time as prior knowledge  
during both training and testing. This prior improves the quality of density reconstructions in our framework. 


The rest of this paper is organized as follows. Section~\ref{sectionproblem} discusses the main dynamic density reconstruction problem. Section~\ref{sectionmethod} presents our proposed method for WGAN-based dynamic density correction. In Section~\ref{sectionexperiments}, we present numerical experimental results and analyses, and finally, we conclude in Section~\ref{sectionconclusion}.

\section{Problem Statement} \label{sectionproblem}
The overall aim of our study is to recover an estimate of 
a time-series of densities $\left \{ \density_t \right \}_{t=t_1}^{t_N}$, with $\density_t \in \mathbb{R}^{N_1 \times N_2 \times N_3}$ (where $N_1$, $N_2$, and $N_3$ denote the spatial dimensions of the 3D densities to be reconstructed), from their corrupted projections or radiographs $\left \{ \mathbf{m}_t \right \}_{t=t_1}^{t_N}$.
The times $t_1, t_2, ..., t_N$ denote $N$ time points at which the radiographic measurements are collected.
The density variations correspond to a 3D object, although, in our studies, we will look at axis-symmetric objects that are fully characterized by the central slice through the (spherical) object. 

The measurements of the density fields correspond to projections or radiographs that are also corrupted by scatter and noise. 
We briefly present the imaging model as follows (see~\cite{mccann_local_2021}).
In the radiographic setup, there is an X-ray source, the object being imaged, and a detector.
The areal density of the object along a ray $r$ connecting the source and detector is denoted
$\rho_{A_i}(r) = \int_{-\infty}^\infty \rho_i(r_x(u), r_y(u), r_z(u))\, \mathrm{d}u$,
where $(r_x(u), r_y(u), r_z(u))$ is a parameterization of the ray $r$, $\rho_i(\cdot)$ denotes a (spatially) continuously varying density (ignoring time) and
$i$ is the object index---
each different material being imaged corresponds to a different $i$.
A simple measurement model with a mono-energetic X-ray source and a single material in the object is that
the number density of photons reaching the detector along ray $r$ is
\begin{equation}
\label{eq:beer}
    I(r) = I_{0}
    \exp \left( - \xi \rho_{A}(r)\right),
\end{equation}
where  $I_0$ is the number density of the incident beam and $\xi$ is the mass attenuation coefficient of the material~\cite{berger_xcom_2010}.
A discrete, finite radiograph is measured in practice representing a finite grid of detectors.
This direct radiograph (without scatter) $\direct$
at each detector pixel is approximated as
\begin{equation} \label{eq:direct-mono}
    \direct[m, n] 
    = \int_{R_{m,n}} I(r)\, \mathrm{d}r
    \approx C I(r_{m,n})
\end{equation}
where $R_{m,n}$ denotes the rays impinging pixel $(m,n)$,
$r_{m,n}$ is the ray through the pixel center,
and
$C$ is a constant (depends on factors such as the detector pixel area).

The final observation is referred to as the total transmission
$\m$
that includes contamination from a \emph{scatter}  signal $\scatter$ and additional noise $\noise$,
i.e., $\m = \direct + \scatter + \noise$.
Considering the entire time-series, the relationship that connects the measurements and the (discretized) density fields 
is the following:
\begin{equation}
    \label{problem_setup}
    \m_t = \direct_t + \scatter_t + \noise_t, \,
    \direct_t = I_0 \exp{\big(-\xi\abel[\density_t]\big)},
\end{equation}
where $\abel$ is the forward operator (e.g., Abel transform), 
$\noise_t$ is typically assumed stochastic (e.g., Poisson or Gaussian noise)~\cite{thibault2006recursive,lu2001noise,demirkaya2001reduction,sauer1991nonstationary}, and $\scatter_t$ is typically assumed to be a (unknown) function of the $\direct_t$ or $\density_t$, e.g., $\scatter_t = \mathbf{k} \ast \direct_t$ obtained as the convolution of the direct radiograph with a scatter kernel $\mathbf{k}$ (e.g., Gaussian kernel or one that depends on the underlying density distribution) is common in the literature~\cite{sun_improved_2010,mccann_local_2021}. The above model can be readily extended to polyenergetic X-ray sources or to objects with multiple materials~\cite{elbakri_statistical_2002,mccann_local_2021}. 
In this work, we focus on single-material objects that are of interest in materials science, inertial confinement fusion, and national security applications, and we work with monoenergetic X-ray sources for simplicity. 

The density time-series reconstruction in~\eqref{problem_setup} is challenging due to the presence of scatter (and noise) that may be imperfectly characterized and due to any mismodeling of the forward operator.
Incorporating priors for the evolution of the underlying density fields over time could provide improved dynamic reconstructions~\cite{myers2011dynamic,bonnet2003dynamic,desbat2007algebraic,jailin2018dynamic,hossain2021high}.
The evolution of the density image over time in a dynamic experiment is governed by physical laws that are described by a system of partial differential equations. 
Utilizing these equations in the reconstruction process would encourage the estimated density time-series to be consistent with the equations of motion that govern their evolution.  
In particular, the density evolution 
is described by ideal Euler equations. The system of PDEs comprise an example continuum model governing density evolution in many realistic scenarios, where dynamic radiography might be applied. In full generality, these partial differential equations (PDEs) may be written in the form~\cite{toro2013riemann}
\begin{gather}
    \partial_t\rho + \nabla\cdot(\mathbf{u}\,\rho) = 0,\quad \rho\,(\partial_t\mathbf{u} + (\mathbf{u}\cdot\nabla)\mathbf{u}) = -\nabla p,\quad \rho\,(\partial_t e  +\mathbf{u}\cdot\nabla e)=-p\,\nabla\cdot \mathbf{u},
\end{gather}
where $\rho$ is the mass density, here, written as a scalar function of spatial coordinates and time, $\mathbf{u}$ is the fluid velocity, $e$ is the specific internal energy, and $p = P(\rho,e)$ is the pressure expressed as a point-wise function of $\rho$ and $e$. The precise form of $P(\rho,e)$ is determined by a fluid's equation of state (EOS), and will differ from material to material. If a non-dissipative fluid's EOS is known, along with appropriate initial and boundary conditions, the Euler equations uniquely prescribe the time evolution of any hydrodynamic variable, including the density.

However, in practice, the EOS and other conditions or parameters for a specific experiment or test case are usually unknown.
So directly using the PDEs to enable dynamic reconstruction is not feasible.
Instead, in our proposed scheme, we use the hydrodynamic models to generate large datasets of density time-series, from which we learn neural network-based priors to enable accurate density estimation as discussed in Section~\ref{sectionmethod}.



\section{Proposed Method} \label{sectionmethod}
In our approach, we assume that a reconstruction of the unknown density fields $\left \{ \density_t \right \}_{t=t_1}^{t_N}$ is available. For example, these could be obtained via a scatter/noise correction and density reconstruction algorithm, in which case, the residual errors from the algorithm would still need to be corrected.
Alternatively, these could be obtained from a traditional reconstruction (e.g., the inverse Abel transform) that ignores the presence of corruptions in the radiograph, thus resulting in distorted density profiles.

We leverage a combined physics-based learning process to train a (3D\footnote{In our framework, the central 2D slices of the azimuth-symmetric 3D density fields are stacked over time to generate the network inputs/targets.}) network as a denoiser, which takes in the noisy density fields denoted as $\left \{ \widetilde{\density}_t \right \}_{t=t_1}^{t_N}$, and outputs estimated clean density fields. 
In our studies, the noisy density fields include the effects of simulated or partially reduced/corrected scatter fields with stochastic variations, and random noise.
The training process (or loss) contains two components: a supervised part which exploits available noisy-clean density time series pairs as training data, and an unsupervised part which, in an adversarial form, aims to regulate the network to reflect the dynamics behind the time series data.
In the following, we first present some preliminaries and then discuss the approach in detail.

\subsection{Review of Wasserstein distance}
Wasserstein distance is a measure of the distance between two probability distributions. In this work, we use the Wasserstein-1 distance. For probability distributions $p_\textrm{real}$ and $p_\textrm{fake}$, the Wasserstein-1 distance between them is defined as \begin{equation}
\label{defn::wasserstein_1_distance}
    W(p_\textrm{real},p_\textrm{fake}) = \inf_{p\in\mathcal{P}(p_\textrm{real},p_\textrm{fake})} \E_{(\x,\y)\sim p} \, \textrm{dist}(\x,\y),
\end{equation}
where $\textrm{dist}:\mathcal{X}\times\mathcal{X}\to \R$ is a distance metric, $\mathcal{P}(p_\textrm{real},p_\textrm{fake})$ is the set of all joint distributions for $x$ and $y$ whose marginal distributions are $p_\textrm{real}$ and $p_\textrm{fake}$ (both defined on a compact vector space $\mathcal{X}$), respectively. In this study, we chose the distance metric in the definition of $W$ to be the standard $\ell_2$ distance.
The Wasserstein-1 distance is also known as the earth-mover's distance. Intuitively, if the distributions are interpreted as two different ways of piling up a certain amount of earth (dirt) over the region $\mathcal{X}$, the earth-moving distance is the minimum cost of turning one pile into the other, where the cost is assumed to be the amount of dirt moved times the distance by which it is moved.

The infimum in the definition of the Wasserstein-1 distance~\eqref{defn::wasserstein_1_distance} is computationally intractable, and thus in order to implement it in numerical experiments, we need a feasible form of this distance. Using Kantorovich-Rubinstein duality~\cite{kantorovich_duality}, the Wasserstein-1 distance can be written as 
\begin{equation}
    W(p_\textrm{real},p_\textrm{fake}) = \sup_{\|f\|_L\le 1} \E_{\x\sim p_\textrm{real}} f(\x) - \E_{\y\sim p_\textrm{fake}} f(\y),
\end{equation}
where the supremum is over all 1-Lipschitz functions $f:\mathcal{X}\to \R$. A function $f$ is 1-Lipschitz if $|f(\x)-f(\y)| \le \|\x-\y\|$. 

\subsection{Adversarial Part}
Through the adversarial component in the training setup, we aim to train the denoiser to output denoised densities (time sequence) that agree with density distributions characterized by the underlying dynamics from simulations. To embed the prior of physical dynamics in the trained model, or in other words, to require that the distribution of denoised density time series is aligned with the distribution of physically valid dynamic time series,
we resort to the adversarial training setting which includes a generator and a discriminator. In this work, the generator (G network) is a 3D-Unet with residual connections~\cite{3DUnet} that serves as the denoiser and the discriminator (D network) is a convolutional neural network which is supposed to tell whether a certain density evolution sequence agrees with the underlying dynamics or not. The residual 3D Unet we use has 4 down-sampling convolutional blocks.

The training objective for the D network is as follows:
\begin{equation}
    \label{D_loss}
    \min_{\Theta_D} \, \underbrace{\E[D(G(\density_{\textrm{noisy}});\Theta_D)] - \E[D(\density_\textrm{clean};\Theta_D)] + \eta\cdot\E[( \|\nabla_{\density}D(\density;\Theta_D)\| -1)^2]}_{\textrm{Loss}_D(\Theta_D)},
\end{equation}
where the last term is the gradient penalty which penalizes the gradient norm of the D network with respect to both real and denoised dynamic time series, thus helping enforce the 1-Lipschitz constraint for the discriminator network, and $\eta \geq 0$ denotes the strength of this regularization term.
The parameters of the discriminator network are denoted $\Theta_D$ and $\density_{\textrm{noisy}}$ and $\density_\textrm{clean}$ denote the noisy and clean density time-series (stacked over time).

\paragraph{Network structure of discriminator} The D network 
has two main parts: convolutional part and a fully-connected part, and ends with a Sigmoid activation function. 
The convolutional part consists of 6 convolutional blocks. Each convolutional block entails three steps, which are respectively a 3D convolution step, a 3D instance normalization step, and a leaky ReLu activation step. The kernel sizes and stride sizes are $(3,4,4)$ and $(1,2,2)$ respectively for the first, second, fourth and sixth convolutional blocks, while for the third and and fifth convolutional blocks, they are $(4,4,4)$ and $(2,2,2)$, respectively. The output of each convolutional block has twice as many channels as its input, except for the first convolutional block whose output channel number is a hyperparameter, which we set as 4. We set the leaky ReLu activation threshold to be 0.2.

The dense part consists of four fully connected layers, each of which contains a linear map followed by a leaky ReLu activation function. The dimension of a dense layer shrinks by 25\% from the previous layer, and the output 
is a scalar. The leaky ReLu activation threshold in the fully-connected blocks is 0.01.


\subsection{Supervised Part}
GAN training has been long known to be unstable due to mode collapse and the sensitivity to hyperparameter settings and balancing of model complexity. Recall the target of our work is to prepare a denoiser that removes stochastic 
perturbations from time series dynamics. 

Motivated by the prior work, we include a normalized squared $\ell_2$ data fidelity term (or normalized mean squared reconstruction error) in the loss function for the G network to direct the early stages of the training and decrease the weight for this term as the training process proceeds. In our study, the normalized $\ell_2$ criterion works better than the normalized $\ell_1$ error in terms of preserving fine local textures (such as discontinuity or sudden jumps of pixel values in the density images) and overall image quality. 
In addition,  a mass-conservation term is included as a regularizer in the training loss as the object mass is preserved over time during the dynamic experiment.
The formulation for learning the G network is thus
\begin{align}
     \label{G_loss}
     \min_{\Theta_G} \, \textrm{Loss}_G(\Theta_G) =  \min_{\Theta_G} \, \lambda\cdot \bigg[&\frac{\|\density_{\textrm{clean}} - G(\density_{\textrm{noisy}};\Theta_G)\|_2}{\|\density_{\textrm{clean}}\|_2}\bigg] \, + \nonumber\\
     &(1-\lambda)\cdot \big( - \textrm{Loss}_D \big) + \lambda_{M}\cdot\E[\|M(G(\density_{\textrm{noisy}};\Theta_G)) - M(\density_{\textrm{clean}})\|_2],
\end{align}
where the parameters of the generator are denoted by $\Theta_G$ and
the term $\E[D(G(\density_{\textrm{noisy}};\Theta_G))]$ in Loss$_D$ has a dependence on the parameters of the G network, $M(\density) = \int \density \,\mathrm{d}V $ is the function that computes the total mass of each density frame in its input and outputs a vector of such masses (over all input frames), $\lambda$ is the weight for the supervised loss term and $\lambda_M$ is the weight for the mass conservation regularizer.

If $\lambda = 1$ throughout the training process, then the training procedure becomes the classic supervised learning process.

\subsection{Post-processing of Denoised Densities}
At test time, the noisy densities are simply passed through the generator in the learned WGAN.
As a further refinement of the denoised densities from the WGAN, we enforce an explicit mass regularization and a total variation prior.
This is similar to fine-tuning of denoised results (e.g., via enforcing data-consistency or regularization) that has been used in other reconstruction networks~\cite{lahiri2022} to bring in additional priors.
We feed the output of the G network into a solver for the following optimization problem:
\begin{equation}
    \label{opt::post-processor}
    \min_{\density}\, \lambda_0 \cdot \frac{\|\density-G(\density_{\textrm{noisy}})\|_2}{\|G(\density_{\textrm{noisy}})\|_2} + \lambda_1 \cdot \underbrace{\|M(\density) - M(\density_{\textrm{clean}})\|_2}_{\textrm{mass fidelity}} + \, \lambda_2 \cdot \underbrace{\|\density\|_\textrm{TV-A}}_{\textrm{smoothness pen.}},
\end{equation}
where $M (\cdot)$ is the same mass computing function as in~\eqref{G_loss}, and $\|\cdot\|_\textrm{TV-A}$ is the anisotropic total variation norm ($\|\mathbf{X}\|_\textrm{TV-A} = \sum_{i=1}^{m-1}\|\mathbf{X}_{i+1,:} - \mathbf{X}_{i,:}\|_1 + \sum_{j=1}^{n-1}\|\mathbf{X}_{:,j+1} - \mathbf{X}_{:,j}\|_1 $ for a matrix $\mathbf{X}\in\R^{m\times n}$). As before, the mass fidelity term in the optimization problem~\eqref{opt::post-processor} assumes the knowledge of true mass of the object (which could be measured prior to the dynamic experiment). The total variation penalty term 
removes spurious details of the density frames while preserving important dynamical details such as edges and shocks and has been a popular regularizer in tomographic image reconstruction. We point out that~\eqref{opt::post-processor} itself can be used as a denoiser for the original noisy density time series and could operate solely based on mass conservation and the conventional regularization (i.e., set $\lambda_0 = 0$).
In that case, we could initialize an iterative solver for~\eqref{opt::post-processor} with the noisy density, which would be cleaned over iterations.

\section{Experiments} \label{sectionexperiments}
We examine the effectiveness of the denoiser trained in the hybrid framework using hydrodynamic simulations. The hybrid training framework contains contributions from supervised loss functions and unsupervised WGAN loss functions. In the following sections, we refer to our training framework as `WGAN-Sup'. Our experiments evaluate the performance of the WGAN-Sup-trained denoiser for removing scatter at various levels as well as removing additional white noise in the radiographs.

\subsection{Experiment settings}
\paragraph{Dataset.}

\begin{figure}
  \centering
  \includegraphics[scale=0.4]{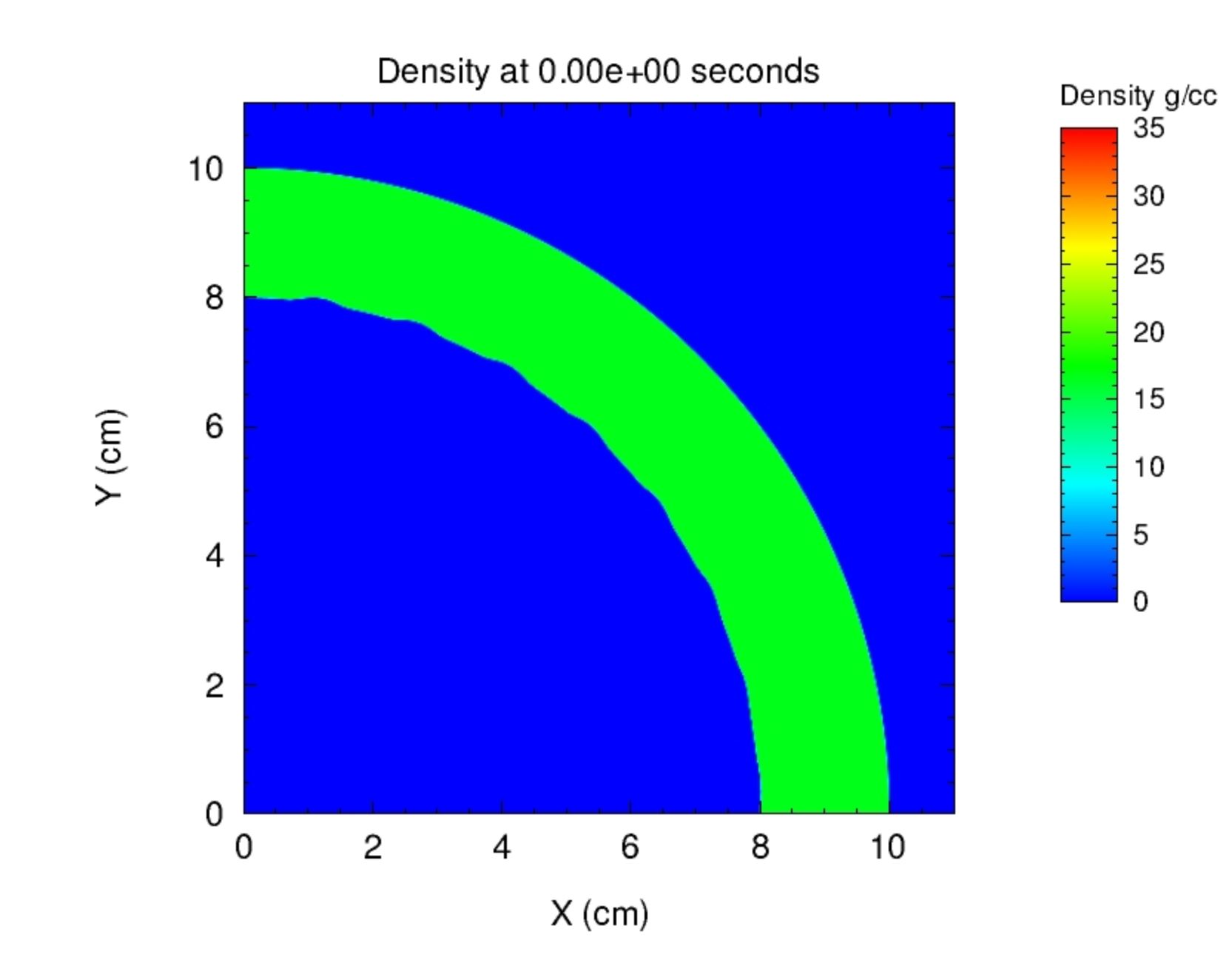}
  \caption{An initial perturbation given to the interior shell in our hydrodynamic data simulation.}
  \label{fig:initialcondition}
\end{figure}

As a test problem, we study a shock propagation in a time-dependent (3D) density profile, created by an implosion of a metallic shell into an air medium.
The study is limited to 
an azimuthal symmetric case in the cylindrical coordinates and Mie-Gr\"uneisen (MG) equation of state.
Simulations are performed with the CTH code~\cite{hertel98a} on a uniform grid with 440 cells covering radius ${R_{D} = 11}$ cm, giving the grid cell size ${\Delta x = 250\, \mathrm{\mu m}}$.
The (spherical) shell is made of Tantalum; its initial density is uniform and equal to that of Tantalum.
The shell initially extends from ${R_{\rm in} = 8}$ cm to ${R_{\rm out} = 10}$ cm of nominal density Tantalum 16.65 gm/cc and is given a uniform implosion velocity of ${v_{\rm impl} = -943}$ m/s (see Figure~\ref{fig:initialcondition}), i.e., it is given a negative constant radial velocity to initiate an implosion.  A perturbation on the interior of the shell is introduced as given by:
$X = R \cos(\theta) + \delta \sin (k \theta)$ and $Y= R \sin(\theta) + \delta \sin (k \theta)$,
where $X$ and $Y$ are the coordinates of the interior shell, $\theta$ is the angle, $R$ is the radius, $\delta$ is the magnitude of the perturbation, and $k$ is the wavenumber of the perturbation.
Figure~\ref{fig:initialcondition} presents an initial perturbation given to the interior shell.  Three different perturbations are examined in the dataset.

\begin{figure}
    \begin{subfigure}{\linewidth}
        \centering
        \includegraphics[width=.8\textwidth]{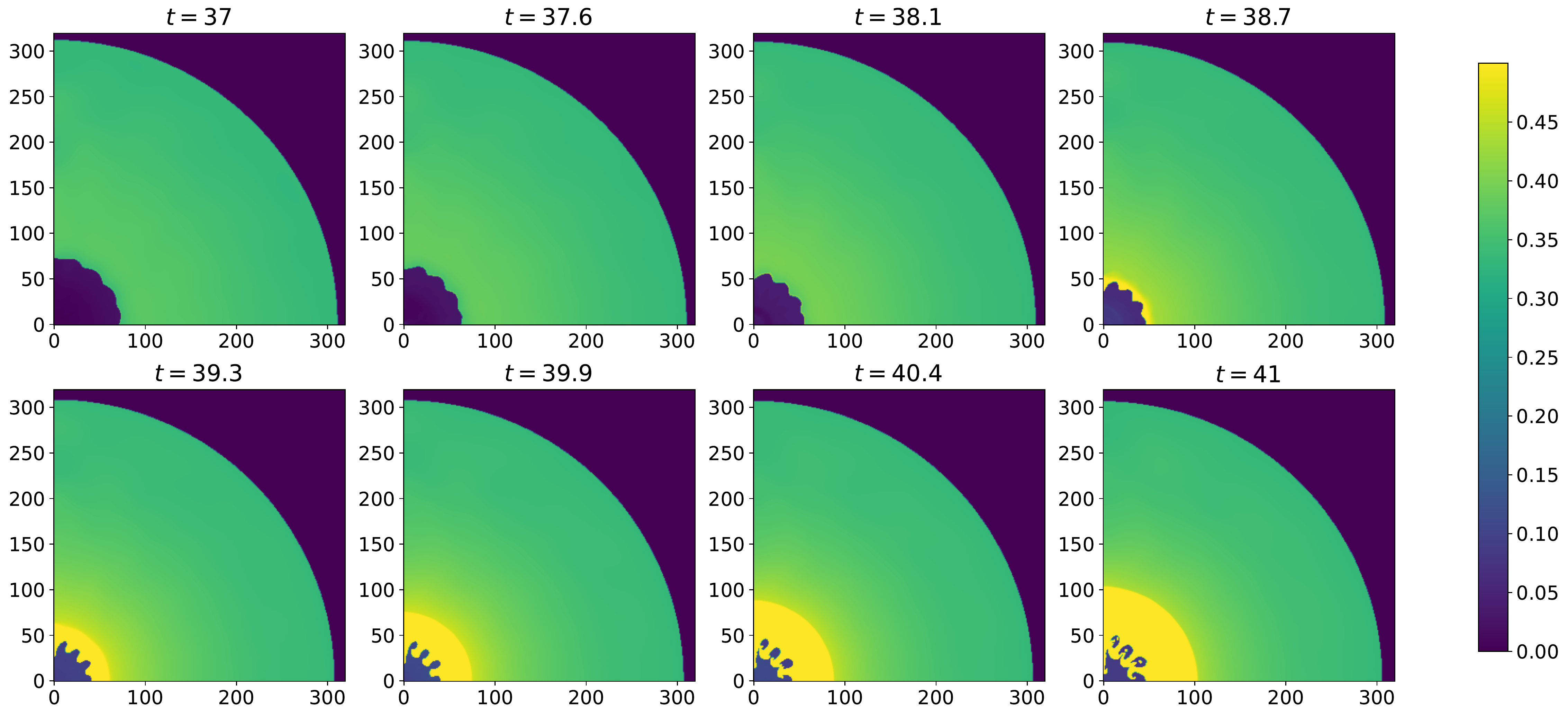}
        \subcaption{}
        \label{fig:dyn_gt_type1}
    \end{subfigure}
    \vfill
    \begin{subfigure}{\linewidth}
        \centering
        \includegraphics[width=.8\textwidth]{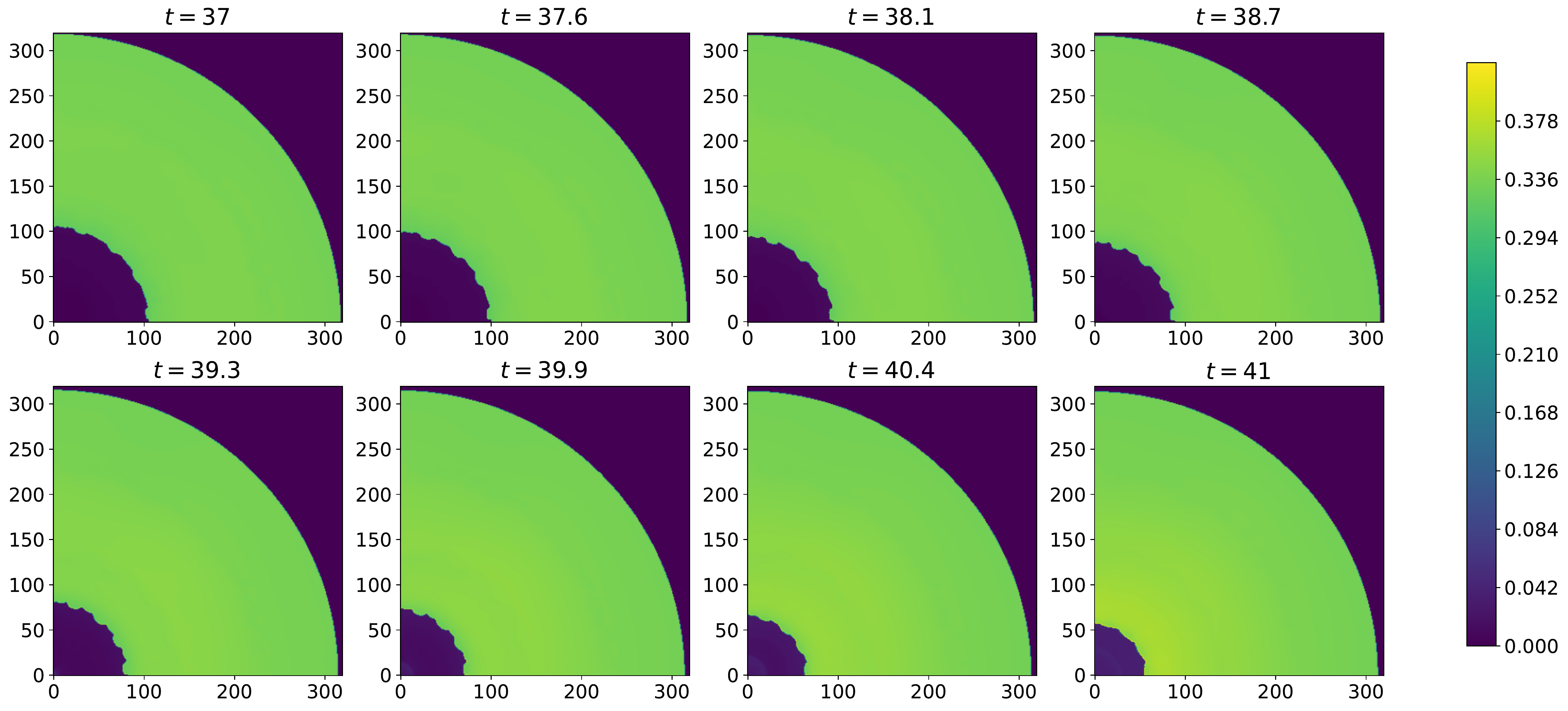}
         \subcaption{}
        \label{fig:dyn_gt_type2}
    \end{subfigure}  
    \caption{Two examples (a) and (b) of density time series from hydrodynamics simulations. }
    \label{fig:dyn_egs}
\end{figure}

Once the Tantalum shell has collapsed, a shock is formed and reflected from the axis.  The shock then interacts with the perturbed inner Tantalum edge. This creates a Richtmyer–Meshkov instability (RMI).  The topology of this interior evolves as depicted in Figure~\ref{fig:dyn_egs}. The expanding shock proceeds to propagate into the
non-constant dynamic density background.



For this test problem, we generate a dataset of simulations covering a 6-dimensional (6D) parameter cube with the parameters characterizing the MG equation of state~\cite{hertel98b}:
\begin{align}
    p(\chi:=1-\frac{\rho_0}{\rho}, T) 
    = \frac{\rho_0 c_s^2\chi\left(1 - \frac12\Gamma_0\chi\right)}
          {(1 - s_1\chi)^2}
    + \Gamma_0\rho_0 c_V (T - T_0),
\end{align}
where $\rho_0$ and $T_0$ are the reference density and temperature, respectively, $c_s$ is the speed of sound, $\Gamma_0$ is the Gr\"uneisen parameter at the reference state, $s_1$ is the slope of the linear shock Hugoniot, and $c_V$ is the specific heat capacity at constant volume.
Out of these parameters, we keep the reference density $\rho_0$ fixed (as stated above) and also keep the reference temperature $T_0$ fixed at 0.0253 eV and vary the parameter set $\{c_s, s_1, \Gamma_0, c_V\}$ as shown in Table~\ref{tab:vary}. In addition, we vary the initial velocity of the shell as indicated in Table~\ref{tab:vary}.

Finally, in hydrodynamic computer codes such as CTH, the stress-tensor components are split into a hydrostatic equation of state and a modified elastic-perfectly plastic constitutive model. In these simulations, we utilize a Preston, Wallace, Tonk (PTW) strength model applicable to metals at high strain rates.  Table~\ref{tab:vary} presents the values of the Mie-Gr\"uneisen (MG) equation of state, Shear Moduli, and initial shell velocities used in the datasets.

\begin{table}[htbp]
  \centering
  \begin{tabular}{cccccc} 
  \hline \hline
    Profiles  & 1   & 2  & 3 & & \\
  \hline
     $v_{\rm impl}$ [m/s]  & 950   & 943.35  & 946.20 & 959.50 & 954.75 \\
  \hline
    $\Gamma_0$  & 1.6   & 1.7  & 1.76 & 1.568 & 1.472 \\
  \hline
    $s_1$  & 1.32   & 1.464  & 1.342 &  & \\
  \hline
    $c_s$ [m/s]  & 339000   & 372900  & 305100 & 355000 &  \\
  \hline
    $c_V$ $[{\rm erg}\;{\rm g}^{-1}\;{\rm eV}^{-1}]$  & $1.6\times 10^{10}$   & $1.76 \times 10^{10}$  & $1.44 \times 10^{10}$ &    &  \\
  \hline
    Shear modulus  & $6.5 \times 10^{11}$   & $6.5 \times 10^{12}$  & $6.5 \times 10^{9}$ &  &  \\
  \hline
  \hline
  \end{tabular}
  \caption{Matrix of parameter values used to develop the simulated dataset. All combinations of above parameters are used to simulate our data.}
  \label{tab:vary}
\end{table}

Altogether, the simulations dataset realizes a parameter cube with $13793$ simulations. Each hydrodynamic simulation data file is comprised of $41$ time frames with the size $320\times320$. We chose frames corresponding to the time instants at $t=37 , 37.6, 38.1, 38.7, 39.3, 39.9, 40.4$, and $41$ (all in milliseconds) from each file as the input time series to train the network in our studies.
Out of all the dynamics simulations, we randomly sample with the same probability $1000$ files for training, $100$ files for validation, and $300$ files for testing.
We normalize all the hydrodynamic time series by a constant normalization factor $50$ to ensure that all inputs to the neural network had approximately the same numerical scale.
Figures~\ref{fig:dyn_gt_type1} and \ref{fig:dyn_gt_type2} are examples of two types of dynamics in the training data.



\paragraph{Scatter and noise generation.}
In our experimental setup, we simulate scatter and noise in the radiographs for each density frame.
The relationship connecting the measured total transmission and the underlying (reference or ground truth) density fields in the time-series is the following:
\begin{equation}
    \m_t = \direct_t + \beta \, \scatter_t + \noise_t = \direct_t + \beta \mathcal{G}[\direct_t] + \noise_t, \direct_t =  \exp{\big( -\xi\abel[\density_t] \big)},
    \label{problem_setup_experiment}
\end{equation}
where $\mathcal{G}[\cdot]$ denotes convolution with a Gaussian filter (implemented using the \texttt{scipy} package, with $\sigma = 2$ for the Gaussian kernel), $\beta$ is a scaling capturing the strength of the scatter field relative to the direct signal, and $\noise_t$ is random (Gaussian) noise.
The choice to model scatter as a kernel convolved with the direct is common in the scatter correction literature~\cite{sun_improved_2010,mccann_local_2021}.
This provides a fast scatter model that is at least representative of models used in practice.
We vary the scatter scaling beta stochastically in a range around a nominal $\beta_0$ (5\% variation on either side) in our simulations.
Finally, we set $I_0=1$ and $\xi = 10^{-2}$ to generate the measurements.


\begin{figure}
    \begin{subfigure}{\linewidth}
        \centering
        \includegraphics[width=.8\textwidth]{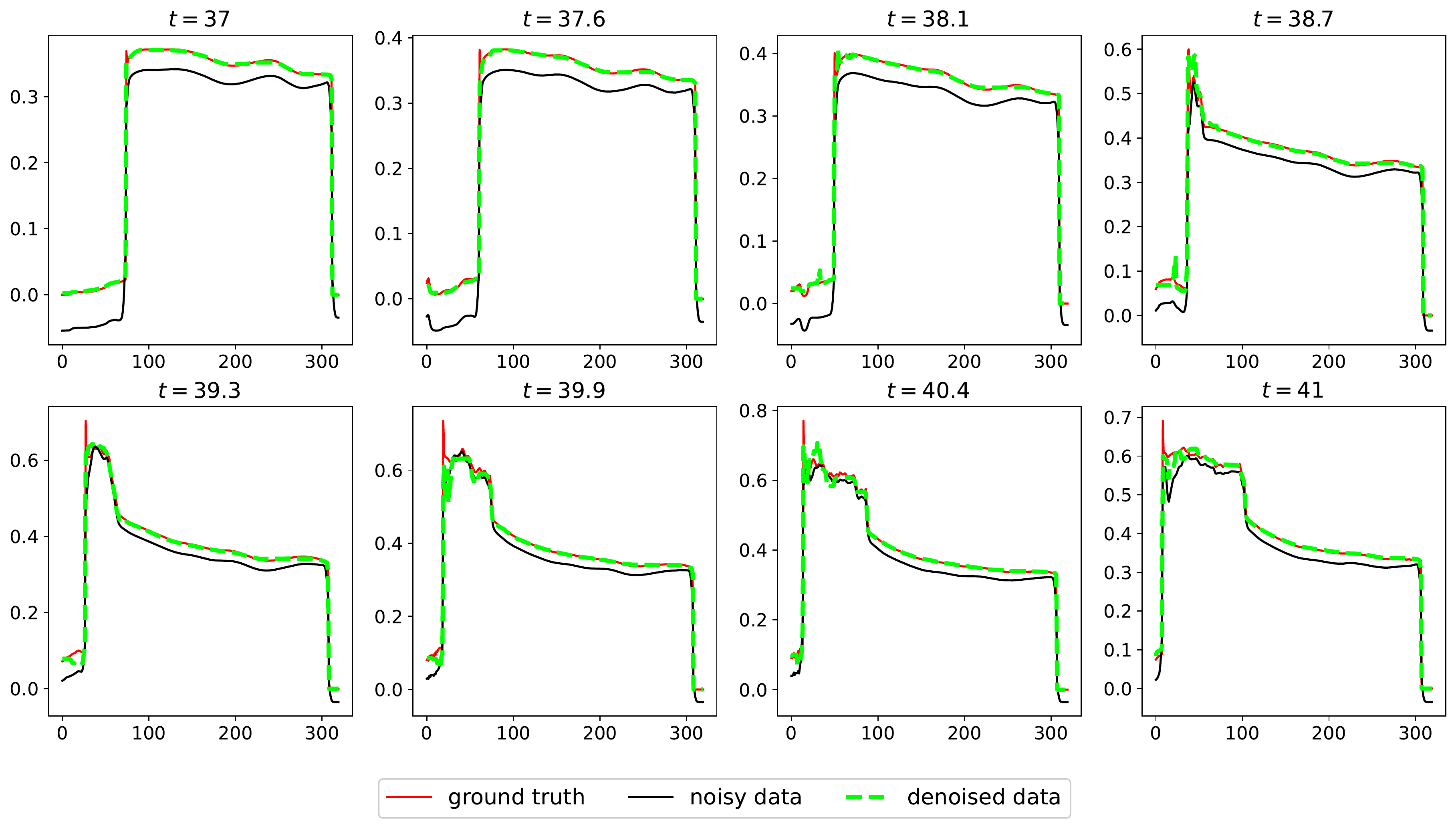}
        \subcaption{Horizontal line profile of Figure \ref{fig:dyn_gt_type1} }
        \label{fig:lineprofile_1}
    \end{subfigure}
    \vfill
    \begin{subfigure}{\linewidth}
        \centering
        \includegraphics[width=.8\textwidth]{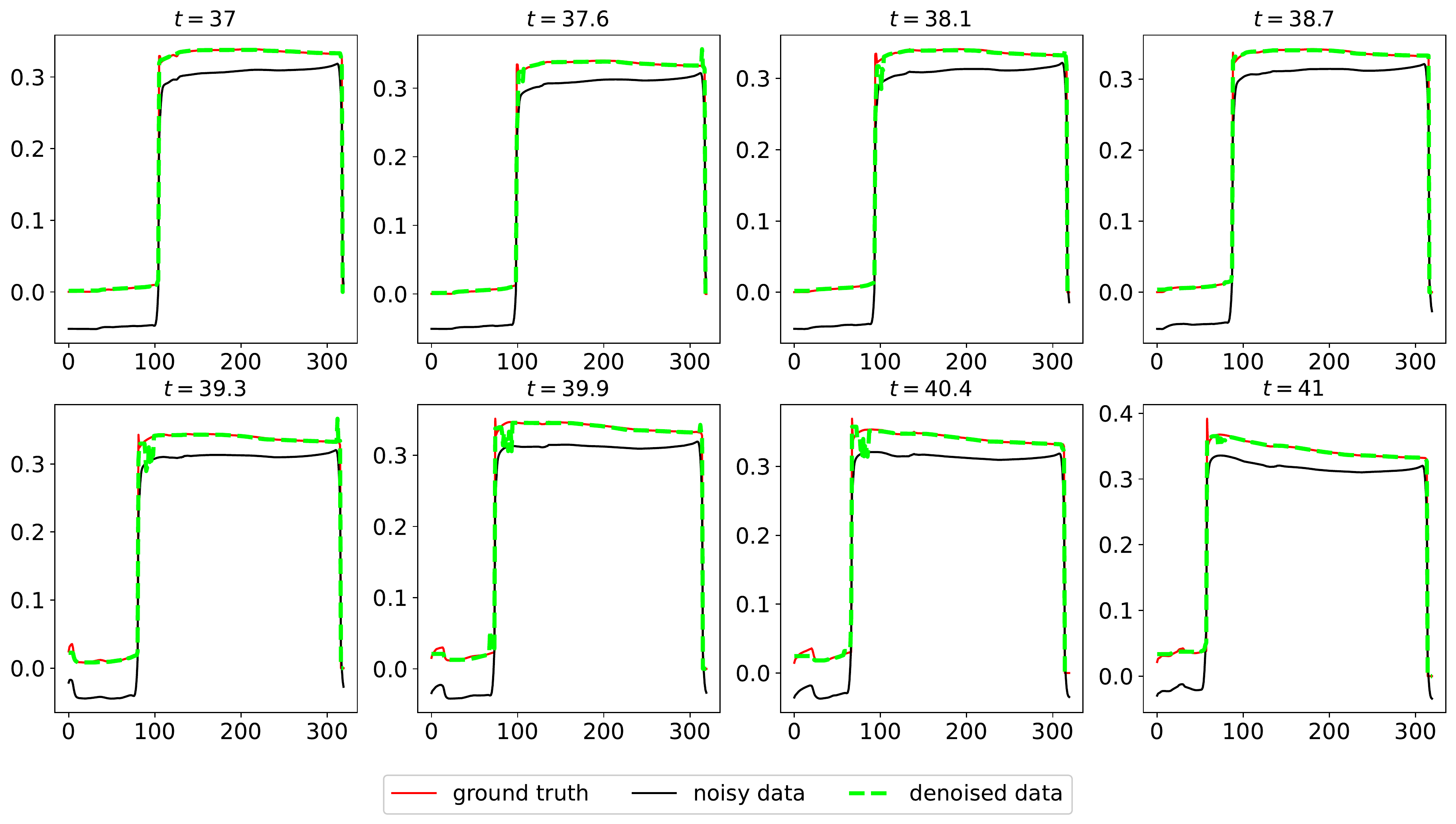}
        \subcaption{Horizontal line profile of Figure \ref{fig:dyn_gt_type2}}
        \label{fig:lineprofile_2}
    \end{subfigure}  
    \caption{Horizontal line profiles corresponding to the example dynamics shown in Figure~\ref{fig:dyn_egs}. Each curve depicts the hydrodynamic information at $y=0$ (horizontal line through center of sphere). The nominal scaling factor $\beta_0$ for the scatter $\scatter$ in~\eqref{problem_setup} is $1$, and there is no white Gaussian noise in the transmission signal. Profiles are shown for the ground truth densities, noisy densities, and densities denoised by our WGAN-Sup framework.
    } 
    \label{fig:line_profiles}
\end{figure}

\paragraph{Hyperparameter settings.} The various hyperparameters for training our networks are carefully chosen as follows. 
The initial weight $\lambda$ for supervised training in the loss function of the G network (for WGAN) is set to 0.99, and decreases by 3\% in every epoch of training. This particular weight is set as a constant 1 to make the training purely supervised (other parameters unchanged) -- a variation, we also investigate. 

The weight for mass fidelity regularizer during training is set as $\lambda_M = 10$. For training the G network in the WGAN, we set the gradient penalty coefficient $\eta=10$, so that the WGAN loss and the gradient penalty contribute roughly equally to the magnitude of the discriminator loss, and both the WGAN component and the Lipschitz constraint can be effectively taken into account during the optimization process. Adam optimizer is used with the momentum parameter $0.9$, mini-batch size $2$ and a learning rate of $2\times10^{-6}$. For training the D network, we use the same optimizer setting except that the learning rate is $10^{-6}$. The update of the G network and D network in the WGAN takes place in an alternating manner, and the update frequency is 1:1. We train both networks for a total of $10$ epochs and select the optimal network based on behavior of the validation loss, which is the normalized $\ell_2$ reconstruction error for the denoised time series. 

\begin{figure}
    \begin{subfigure}{\linewidth}
        \centering
        \includegraphics[width=\linewidth]{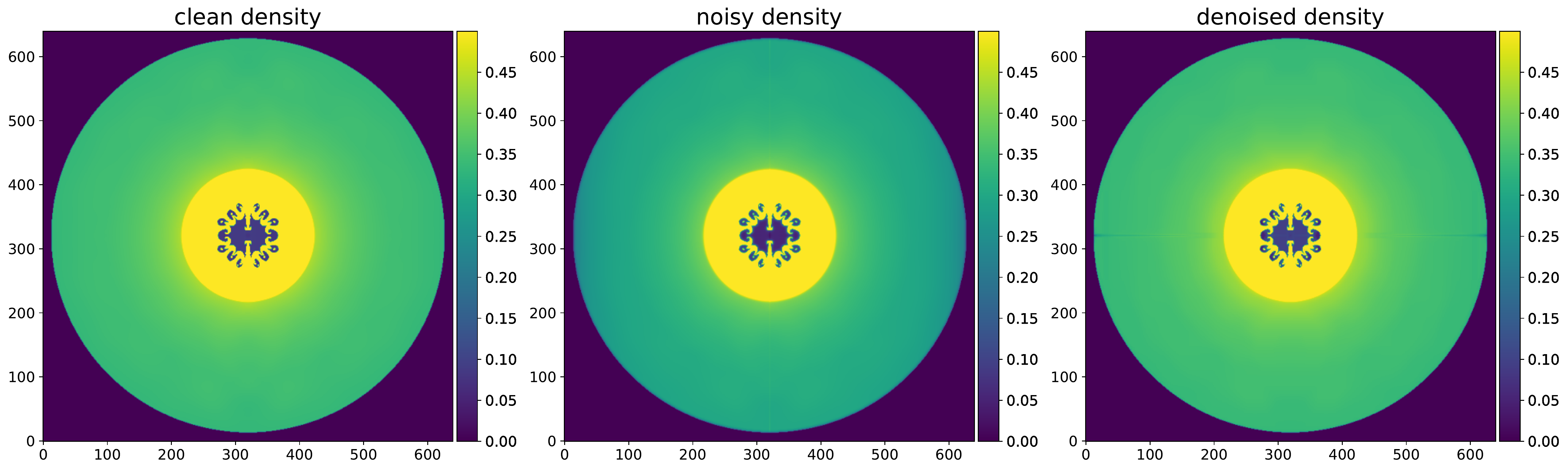}
        \subcaption{}
        \label{fig:density_cmp_type1}
    \end{subfigure}
    \vfill
    \begin{subfigure}{\linewidth}
        \centering
        \includegraphics[width=\linewidth]{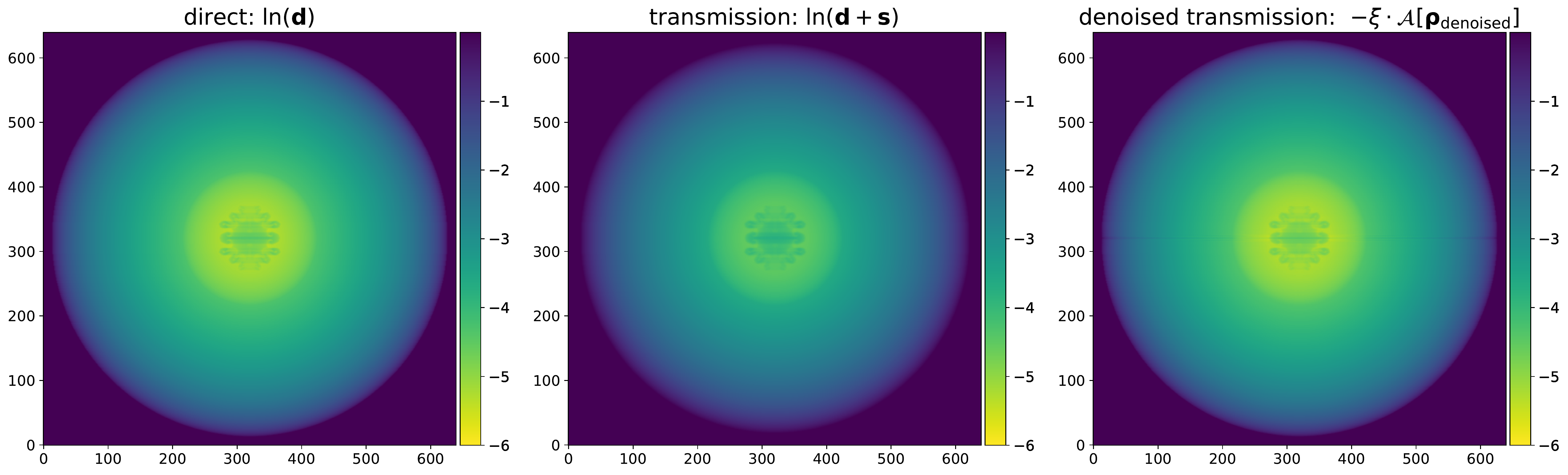}
        \subcaption{}
        \label{fig:radiograph_cmp_type1}
    \end{subfigure}  
    \caption{Density images and radiographs at $t=41$ of the dynamic time series in Fig.~\ref{fig:dyn_gt_type1}.
    (a) Comparison between the clean density frame, noisy density frame (whose radiograph is corrupted by scatter), and denoised density frame.
    (b) Comparison between the corresponding clean radiograph frame, noisy radiograph (with scatter), and denoised radiograph frame. Radiographs are shown in log scale.}
    \label{fig:frame_cmp_type1}
\end{figure}

For the post-processing step for refining the network output, the weight for the data fidelity term is set as $\lambda_0 = 5$, the weight for the total variation term is $\lambda_2 = 10^{-4}$, and the weight for the mass conservation term is $\lambda_1 = 100$. The RMSprop solver is used to solve the optimization problem~\eqref{opt::post-processor} with a learning rate of $10^{-5}$ and maximum iteration count of $7000$.

\paragraph{Baselines.}
We compare the proposed WGAN-Sup framework for density time series correction with variations of the proposed setup as well as simple baselines.
Given a corrupted radiograph (of axis-symmetric object) and assuming a mono-energetic X-ray source, an inverse Abel transform of the radiograph would be the simplest baseline.
This approach does not remove the underlying scatter or noise, and is in fact the noisy density input to the networks in our framework.
Another baseline method for denoising the noisy density
relies on the mass conservation condition (assuming object mass is known) and standard (TV) regularization, which corresponds to solving the optimization problem~\eqref{opt::post-processor} with $\lambda_0 = 0$. The weights for the mass fidelity term and anisotropic total variation term in this case are set as $10^2$ and $10^{-4}$, respectively. The RMSprop solver is used for this optimization process with a learning rate of $10^{-5}$ and maximum iteration count $10^4$, and the optimization is initialized with the noisy densities.

\subsection{Results}
We will rely on two variations of our framework to train our network-based denoisers: WGAN-Sup training and (purely) supervised training. In this section we show their respective performances with respect to denoising the scatter from hydrodynamic time series.


\begin{figure}
    \begin{subfigure}{\linewidth}
        \centering
        \includegraphics[width=\linewidth]{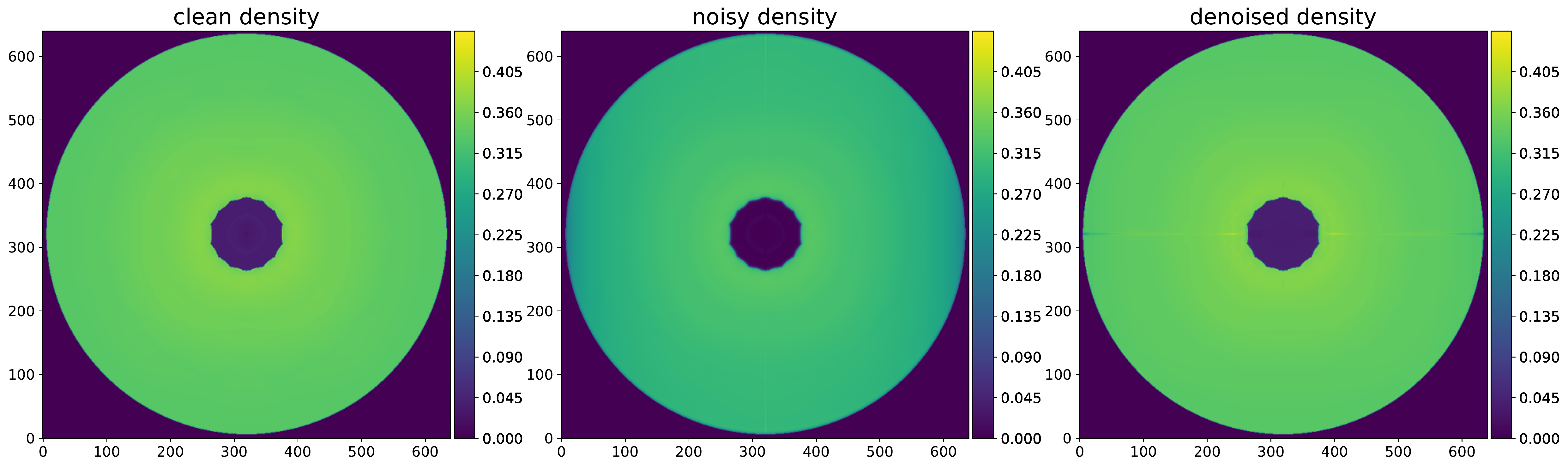}
        \subcaption{}
        \label{fig:density_cmp_type2}
    \end{subfigure}
    \vfill
    \begin{subfigure}{\linewidth}
        \centering
        \includegraphics[width=\linewidth]{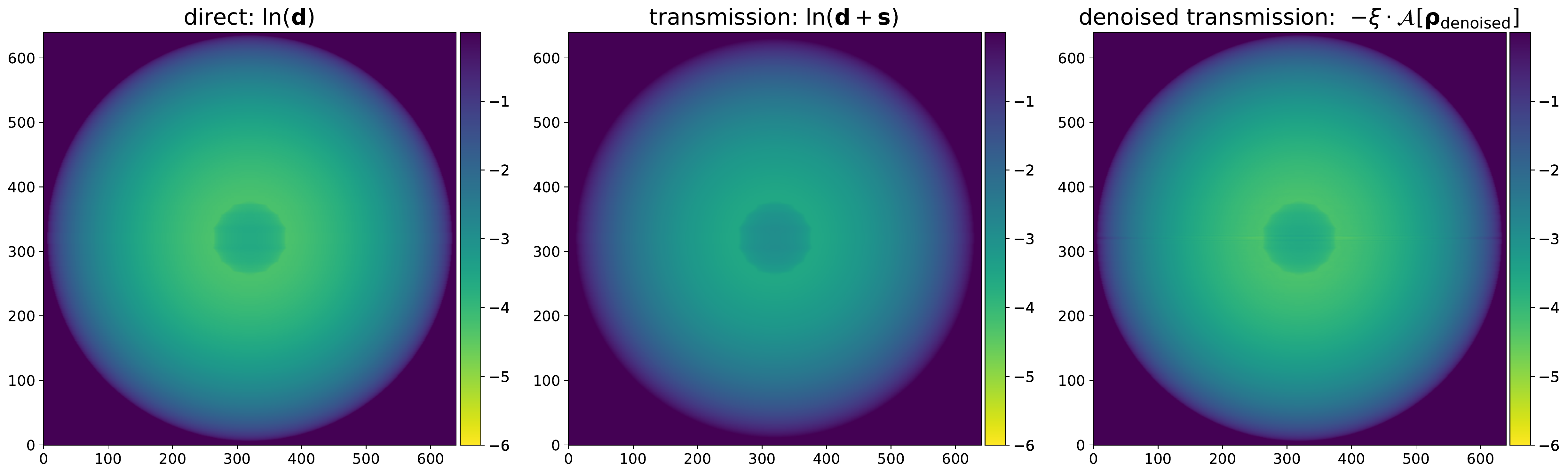}
        \subcaption{}
        \label{fig:radiograph_cmp_type2}
    \end{subfigure}  
    \caption{Examples of clean, noisy, and denoised density and radiograph frames at $t=41$ corresponding to the time series in Figure~\ref{fig:dyn_egs}. The density frames are symmetric with respect to the coordinate axes. (a) Comparison between the clean density frame, noisy density frame (obtained from radiograph corrupted by scatter), and denoised density frame. (b) Comparison between the corresponding clean radiograph frame, noisy radiograph (with scatter), and denoised radiograph frame. Radiographs are shown in log scale.}
    \label{fig:frame_cmp_type2}
\end{figure}

\begin{figure}
    \centering
    \includegraphics[width=\linewidth]{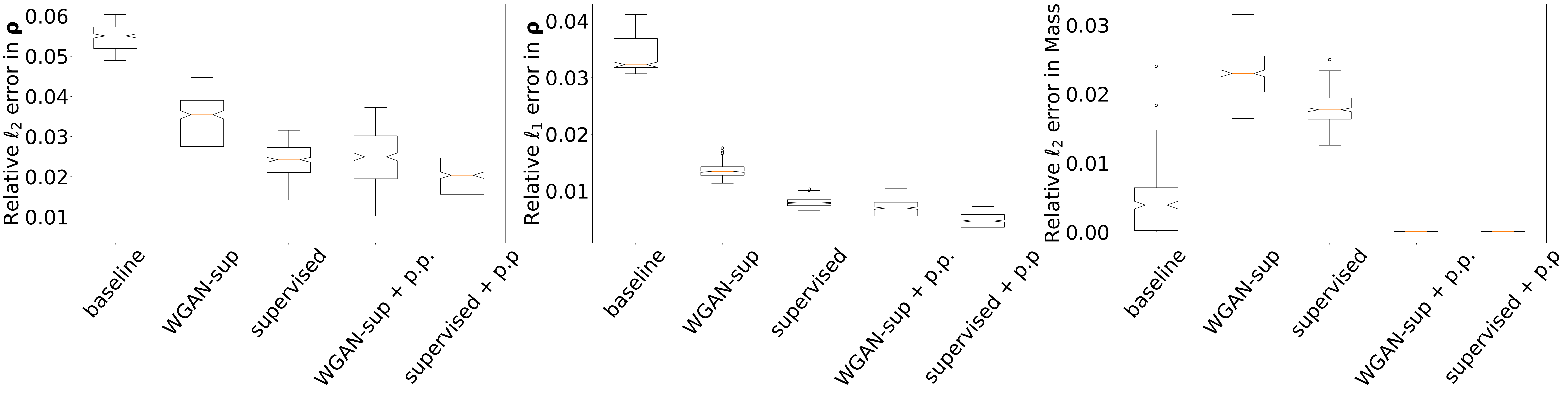}
    \caption{Comparison of overall denoising error in both density (left and middle) and mass (right) between the baseline or classical denoising algorithm and the proposed methods when the nominal scaling of the scatter is $\beta_0 = 1$. The range of the normalized $\ell_2$ error for the noisy or undenoised time series in the testing dataset is $[0.13, 0.16]$ and the corresponding range for the normalized $\ell_1$ error is $[0.12, 0.15]$. The baseline method uses the optimizer for~\eqref{opt::post-processor} with $\lambda_0=0$\protect\footnotemark.
    The proposed methods include the WGAN-Sup-trained denoiser, the WGAN-Sup-trained denoiser plus the post-processing (p.p.) step, the (purely) supervised-trained denoiser, and the supervised-trained denoiser plus the post-processing (p.p.) step. The relative $\ell_2$ error of mass is defined as $\frac{|M(\density_{\textrm{denoised}}) - M(\density_{\textrm{clean}})|}{M(\density_{\textrm{clean}})}$ w.r.t.\ one frame and the presented values are the average over 8 frames in a time series instance.}
    \label{fig:err_gaussian}
\end{figure}

Figure~\ref{fig:err_gaussian} shows the density denoising (only scatter and no Gaussian noise) performance comparison (using box plots) between the classic denoiser that only exploits the mass conservation condition and TV regularization, and the WGAN-Sup or purely supervised trained denoisers. 
The denoisers involving learned networks all outperform the classical method and provide significantly lower errors than the original noisy densities.
Moreover, the additional post-processing step and regularization leads to lower errors generally than the networks' outputs.
We point out that although the mass conservation-based optimization can give a smooth denoising output, it is unable to recover fine local structures in the density frames. Specifically, Figure~\ref{fig:line_profiles} shows the agreement between the WGAN-Sup denoised line profiles (at the location $y=0$) and the ground truth line profiles. The horizontal denoised line profiles from the mass conservation-based optimization program are smooth and maintain the correct shape, but do not differ from the noisy line profiles much.

Figures~\ref{fig:frame_cmp_type1} and~\ref{fig:frame_cmp_type2} show the clean, noisy, and WGAN-Sup network-denoised densities and corresponding radiographs for two cases. The denoised images clearly look close to the ground truth or clean images.

\begin{figure}
    \begin{subfigure}{\linewidth}
        \centering
        \includegraphics[width=.8\linewidth]{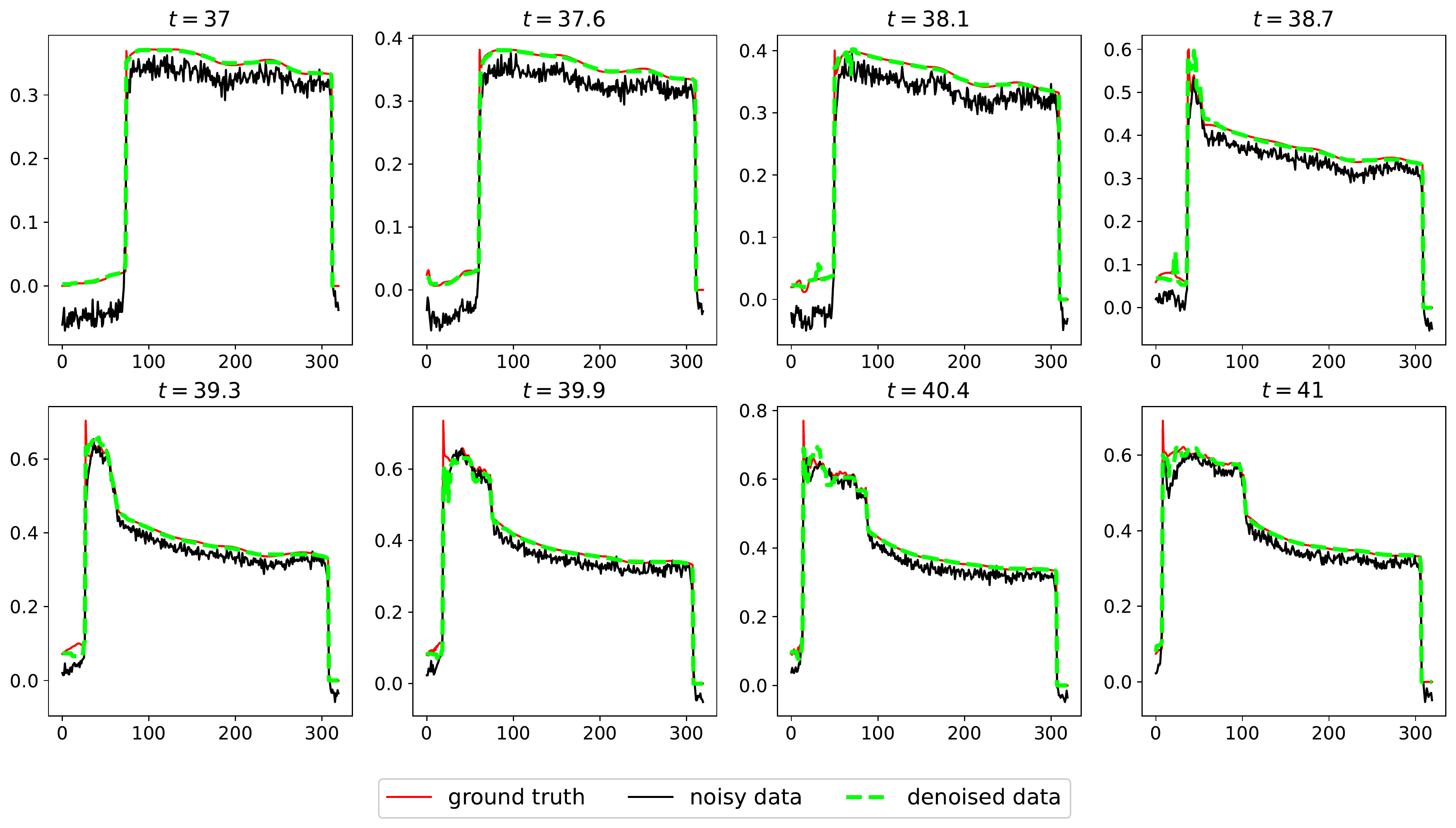}
        \subcaption{Horizontal line profile of Figure \ref{fig:dyn_gt_type1} }
        \label{fig:lineprofile_whitenoise_1}
    \end{subfigure}
    \vfill
    \begin{subfigure}{\linewidth}
        \centering
        \includegraphics[width=.8\linewidth]{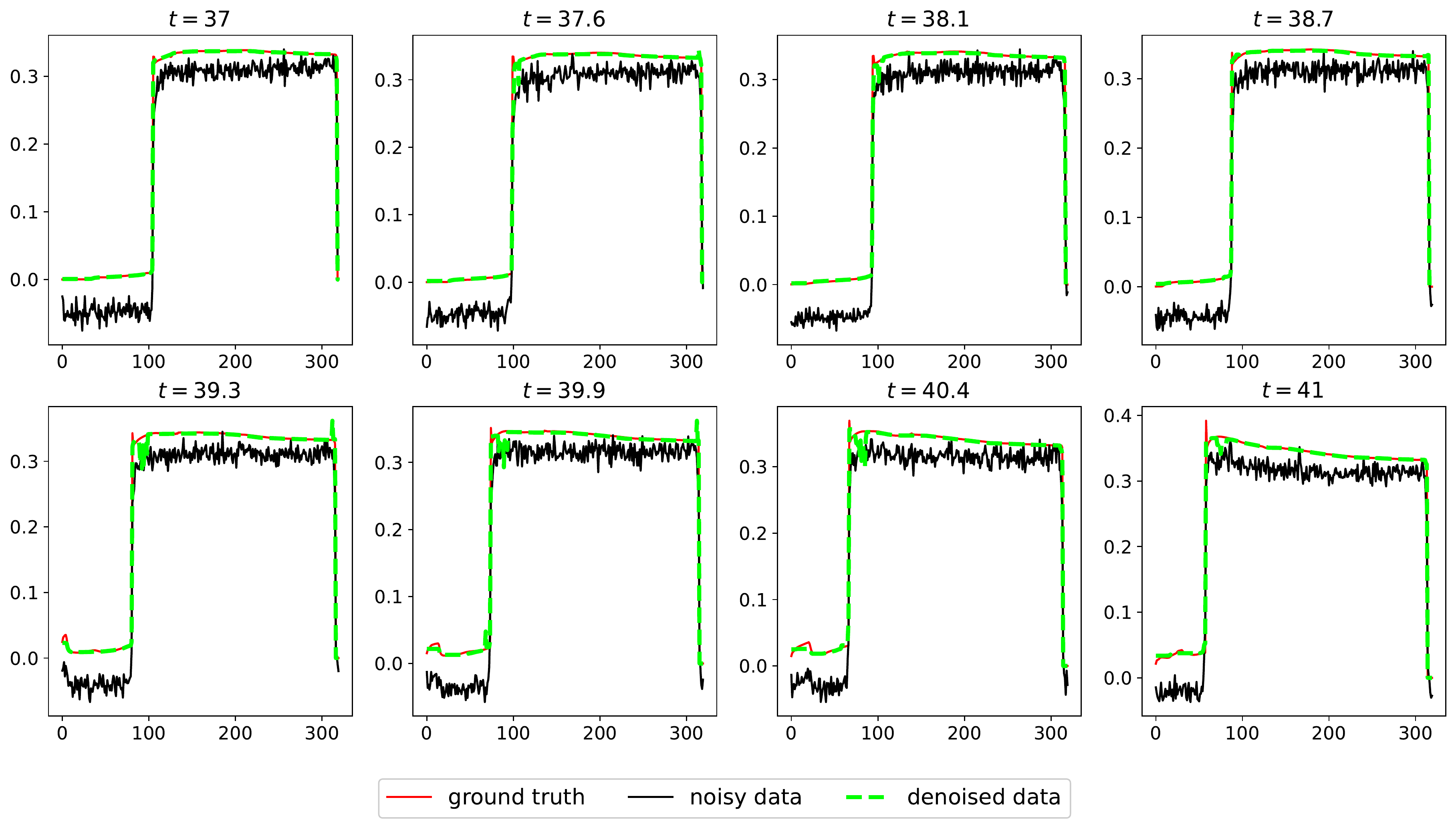}
        \subcaption{Horizontal line profile of Figure \ref{fig:dyn_gt_type2}}
        \label{fig:lineprofile_whitenoise_2}
    \end{subfigure}  
    \caption{Performance with additive noise. Horizontal line profiles corresponding to the example dynamics shown in Figure~\ref{fig:dyn_egs}. Each curve depicts the hydrodynamic information at $y=0$ (horizontal line through center of sphere). The nominal scaling factor $\beta_0$ for the scatter $\scatter$ in~\eqref{problem_setup_experiment} is 1. The transmission signal contains Gaussian noise at each time point, i.e., $\m_t = \direct_t + \mathcal{G}[\direct_t] + \noise_t$, where $\noise_t \sim \mathcal{N}(0,10^{-4}\mathbf{I})$, with $\mathbf{I}$ being the identity matrix. }
    \label{fig:line_profiles_whitenoise}
\end{figure}

\paragraph{Effect of both scatter and random additive noise.}
In the aforementioned cases, we have primarily studied the removal of various levels of scatter in noisy density time-series.
Here, we consider a variant of that experiment setting, where the noise in the radiograph not only comes from scatter but also includes white noise. Figure~\ref{fig:line_profiles_whitenoise} shows the horizontal line profiles for the same dynamic time series as in Figure~\ref{fig:dyn_egs}, however, the noisy densities are corrupted by both scatter and random noise from the transmission domain.
We observe that the WGAN-Sup trained denoiser (trained observing both random scatter and random Gaussian noise) can still perform well for removing noise and its results show fine local line profile structures. 

\begin{figure}
    \centering
    \includegraphics[width=\textwidth]{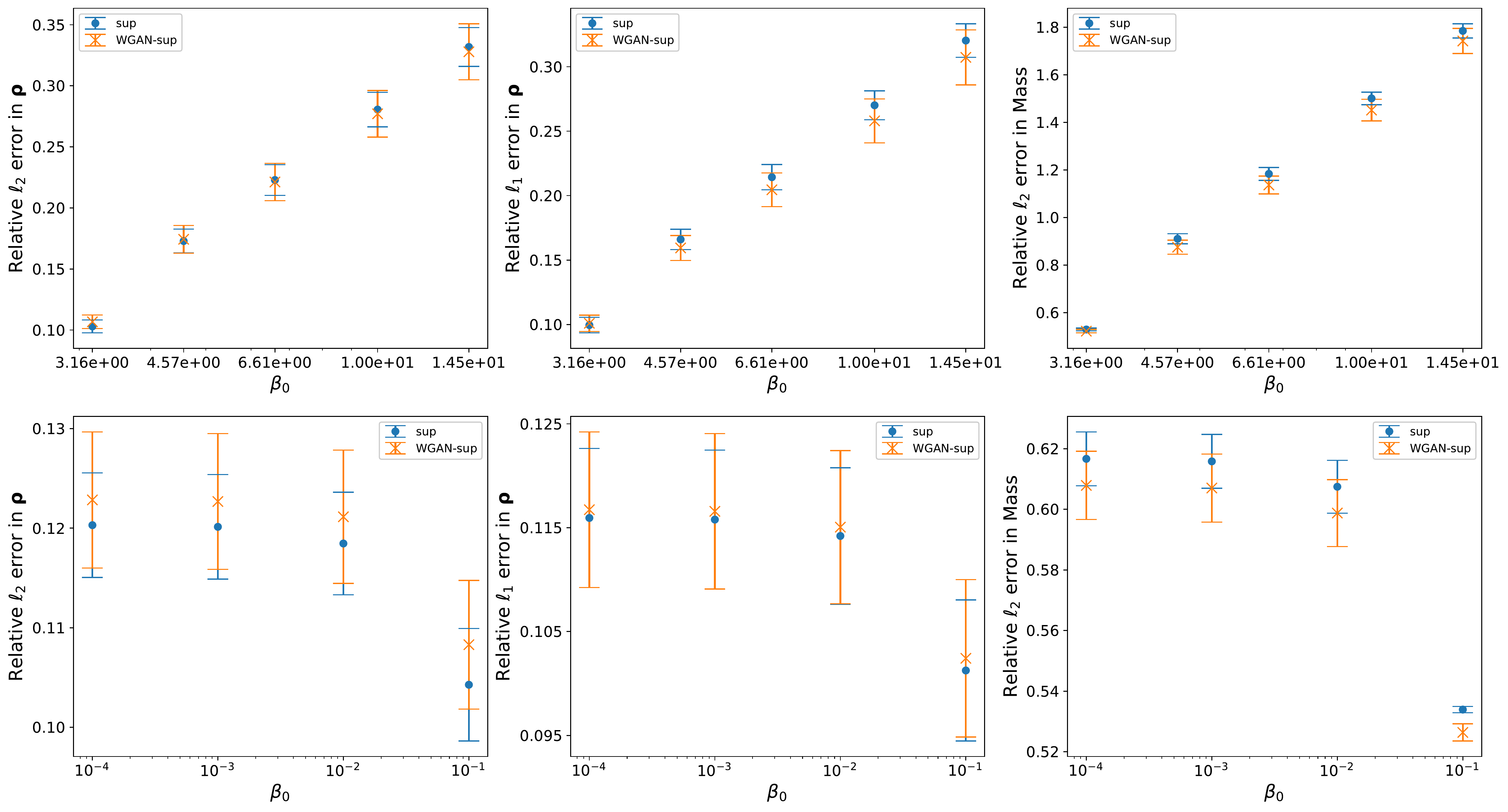}
    \caption{Generalization to different scatter levels. The nominal scaling of the scatter, i.e., $\beta_0$, takes values in two ranges: from $10^{-4}$ to $10^{-1}$, and from $10^{0.5}$ to $10^{1.2}$.
    Here, we use the networks trained with the nominal $\beta_0 = 1$.
    The relative $\ell_2$ error of mass is defined as $\frac{|M(\density_{\textrm{denoised}}) - M(\density_{\textrm{clean}})|}{M(\density_{\textrm{clean}})}$ w.r.t.\ one frame and the presented values are the average over 8 frames in a time series instance. The distance between the central marker and its corresponding whisker indicates one standard deviation.}
    \label{fig:err_abel_gaussian_generalize}
\end{figure}

\begin{figure}
    \centering
    \includegraphics[width=\textwidth]{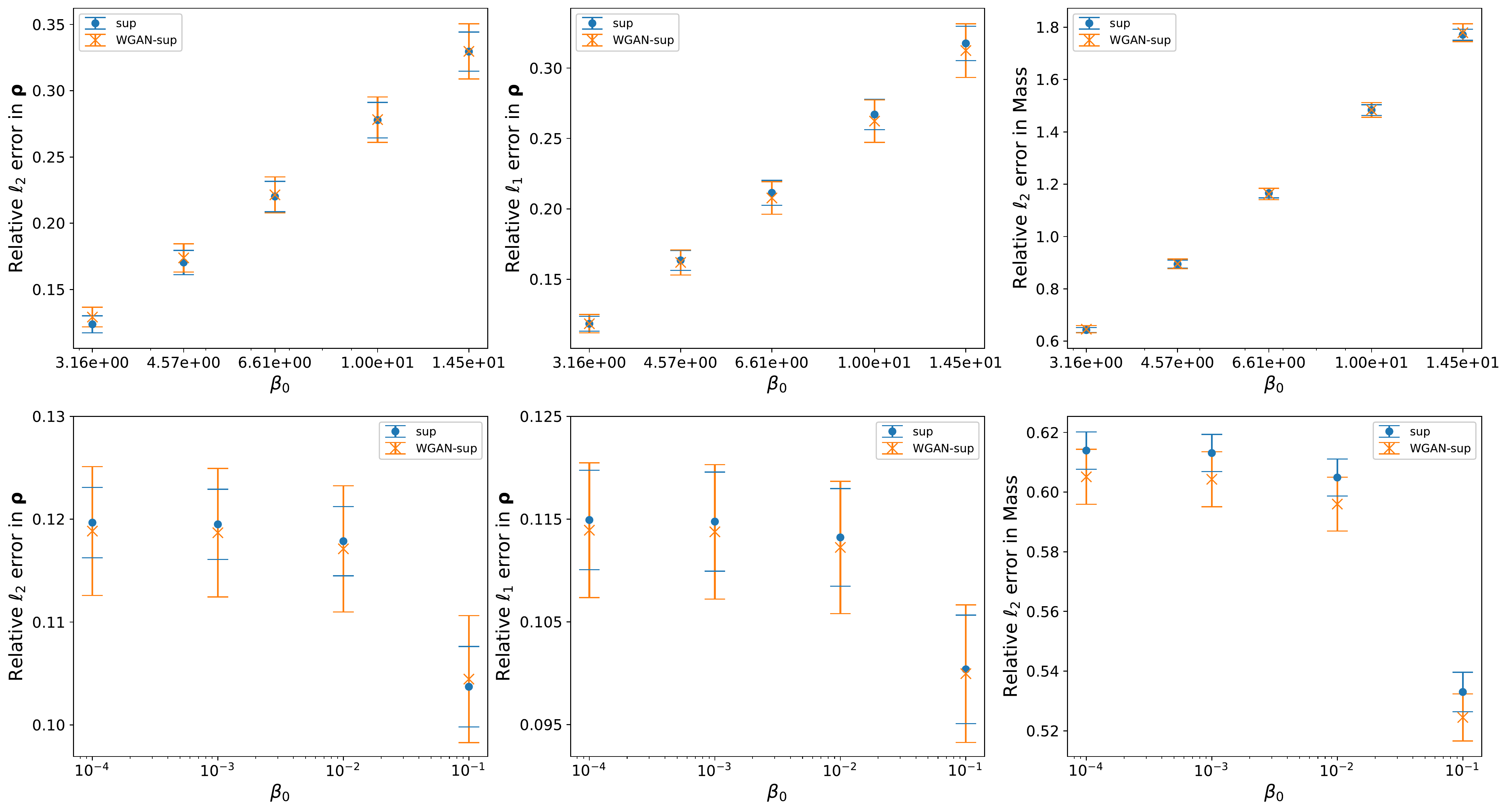}
    \caption{Generalization to different scatter levels in the presence of additive Gaussian noise. The nominal scaling of the scatter, i.e., $\beta_0$, takes values in two ranges: from $10^{-4}$ to $10^{-1}$, and from $10^{0.5}$ to $10^{1.2}$.
    Here, we use the networks trained with nominal $\beta_0 = 1$.
    The relative $\ell_2$ error of mass is defined as $\frac{|M(\density_{\textrm{denoised}}) - M(\density_{\textrm{clean}})|}{M(\density_{\textrm{clean}})}$ w.r.t.\ one frame and the presented values are the average over 8 frames in a time series instance. The distance between the central marker and its corresponding whisker indicates one standard deviation.}
    \label{fig:err_abel_gaussian_double_generalize}
\end{figure}

\paragraph{Generalization to other artifact settings.}
To study the generalization power of the learned networks (supervised and WGAN-sup),
we apply the networks trained with the nominal scatter scaling of $\beta_0 = 1$ to the test cases, but with different scatter ratio and additive noise combinations at test time. 
We consider two noise settings: the noise in the measurements is due to scatter $\scatter_t$ only (sole scattering setting, Figure~\ref{fig:err_abel_gaussian_generalize}), and the noise is due to both scatter $\scatter_t$ and additive white noise $\noise_t$ (Figure \ref{fig:err_abel_gaussian_double_generalize}). The scattering scaling factor $\beta_0$ varies in two ranges: from $10^{-4}$ to $10^{-1}$, which simulates cases where the scatter level is less than that 
in the training set,
and from $10^{0.5}$ to $10^{1.2}$, which corresponds to larger scatter levels than in the training set. 
The WGAN-Sup denoiser shows consistently better performance in both noise
settings in preserving the true mass of the dynamic objects across a variety of scatter levels.
In terms of denoising accuracy, the WGAN-Sup denoiser demonstrates advantages when the scatter scaling $\beta_0>1$ in both noise settings. When $\beta_0<1$, the WGAN-sup denoiser performs better than the pure supervised denoiser in the scatter + additive white noise setting while worse in the sole scattering setting.

\section{Conclusions} \label{sectionconclusion}
This work presents an approach for dynamic quantitative density reconstruction from corrupted radiographs by learning a denoiser by WGAN-sup training framework from large hydrodynamics simulations.
The generator in the WGAN part is trained to denoise corrupted density time series data so that its  output distribution matches the underlying (physically viable) hydrodynamics.
We include supervision during the GAN training for stability and
incorporate additional priors and physics-based regularizers (mass conservation) during training or for post-processing the densities cleaned by the network.
We simulate full or partial scatter with stochastic variations as well as random noise during training and testing.
The proposed framework significantly outperforms conventional radiographic reconstruction and non-learning based regularized denoising, and provides density time-profiles that accurately matched the underlying dynamics and physical properties.

\paragraph{Potential future directions.} Future work could consider
incorporating properties and symmetries of the underlying (but not fully known/characterized) PDEs in our machine learning framework and building a full reconstruction pipeline with joint forward and scatter modeling, density reconstruction, and physics/hydrodynamics priors. 
It is also interesting to investigate how well the network denoising approach can handle or be trained to handle noise coming from prior scatter correction and density reconstruction algorithms when they are applied to dynamic data with improperly characterized scatter or other corruptions.
Another potential direction is to exploit a loss function term that characterizes the fidelity of the reconstruction result with respect to particular signal features. In the hydrodynamic setting, shocks and edges at which discontinuities arise are of specific interest, and thus the accuracy of the denoised result in the region where such discontinuities appear should have more weight during training. Our investigation with the relative $\ell_1$ norm and/or the total variation penalty suggest that there is still space for crafting a better error characterization to highlight the data fidelity in regions where the dynamics demonstrate more volatility.

\paragraph{Acknowledgments}
Zhishen Huang and Saiprasad Ravishankar thank Siddhant Gautam for helpful discussions.

\paragraph{Disclosures}
The authors declare no conflicts of interest.

\paragraph{Data Availability Statement}
The code for the proposed algorithms is available at \url{https://github.com/zhishenhuang/hydro}. Data underlying the results presented in this paper are not publicly available at this time but may be obtained from the authors upon reasonable request.

\bibliographystyle{alpha}
\bibliography{sample,refs_mikepaper}

\end{document}